# Direct observation of phase transitions in Archimedean truncated tetrahedrons under quasi-2D confinement

David Doan[1]†, John Kulikowski[1]†, X. Wendy Gu[1]*

[1]Department of Mechanical Engineering, Stanford University; Stanford, California, 94305, USA.

* Corresponding author. Email: xwgu@stanford.edu

† These authors contributed equally to this work

Colloidal crystals are used to understand fundamentals of atomic rearrangements in condensed matter and build complex metamaterials with unique functionalities. Simulations predict a multitude of self-assembled crystal structures from anisotropic colloids, but these shapes have been challenging to fabricate. Here, we use two-photon lithography to fabricate Archimedean truncated tetrahedrons and self-assemble them under quasi-2D confinement. Under a small gravitational potential, these particles self-assemble into a hexatic phase, which has not yet been observed or reported for this shape. Under additional gravitational potential, the hexatic phase transitions into a quasi-diamond two-unit basis. In-situ imaging reveal this phase transition is initiated by an out-of-plane rotation of a particle at a crystalline defect and causes a chain reaction of neighboring particle rotations. Our results provide a framework of studying different structures from hard-particle self-assembly and demonstrates the ability to use confinement to induce unusual phases.

**Keywords:** *hard particle, self-assembly, colloids*

Colloidal particles can self-assemble into ordered crystals with extraordinary nano and mesoscale complexity[1] and unique optical, electronic, and magnetic properties[2,3]. These emergent properties depend on the properties of the constituent particle and the crystal phase of the final ordered structures. The phase behavior of self-assembled colloidal structures depend on a variety of factors, such as shape, surface interactions, and external fields[1].

Two-dimensional (2D), hard-particle colloidal systems are of interest because they are entropically driven, and their final assembly state solely depends on the shape and packing fraction of the particles. Previous computational[4] and experimental studies have shown interesting crystallization behavior (from liquid to solid) and crystal structures in 2D systems consisting of spherical colloids[5,6], ellipses[7,8], rods and rectangles[9–11], squares[12,13], triangles[14,15], and hexagons[16]. Three-dimensional (3D) self-assembled structures have also been experimentally attained, but usually occur under complex interactions[17–21]. Hard-particle 3D assemblies have been extensively predicted in simulation[22], but are challenging to experimentally achieve and image.

Colloidal crystals are often described as programmable materials[23] but typically form static structures that cannot be reconfigured into different crystals once assembled, or can only be dissembled and re-assembled into the same structure[24]. The ability to directly switch between distinct crystal structures is analogous to solid-solid phase transitions in atomic matter and has previously been studied in different colloidal self-assembled systems[25]. Phase transition kinetics in soft spherical colloids have been previous studied under electric fields[26]. Phase transitions that maintain crystal symmetry can be induced in DNA-functionalized nanoparticle superlattices by inserting additional nanoparticles or DNA linkers[27,28]. Phase transitions have also been investigated with hard-particle spherical colloids[29,30], but in general, these types of studies have been limited to 2D systems that require complicated external fields. In hard-particle systems, one strategy is to change the colloidal shape to an anisotropic, or higher order polygon, which can result in complex phase behavior such as crystal-crystal or solid-solid phase transitions. Colloidal squares[13] that were assembled in 2D have shown complex phase behavior as a function of packing fraction. Superballs[31] that have been assembled in 3D have also shown solid-solid phase transition under different osmotic pressures. However, these superball assemblies show similar phase behavior as those found in 2D systems[12,13]. In addition to shape change, an external potential, such as confinement or boundary conditions, can also play a large role in the possible accessible crystal phases[32–35].

In general, these previous hard-particle phase transitions lack complex phase transformations, such as those in which the "atoms" change coordination number, or transform between crystal lattice systems. Further progress in this field of work could lead to metamaterials with rapidly switchable properties and functional structures. Elucidating the kinetics of colloidal phase transitions could also provide understanding of solid-solid phase transitions in atomic solids, which remain controversial even for elemental materials due to the challenges of observing dynamic behavior at the atomic scale[36,37]. The advantage of these colloidal systems is that they can be imaged at a spatial and temporal resolution that cannot be achieved in real atomic systems, even with state-of-the-art experimental tools such as transmission electron microscopy[38] or ultrafast X-ray diffraction[39–41].

In this work, we assemble lithographed Archimedean truncated tetrahedrons (ATT) at an interface to achieve quasi-2D confinement. This strategy takes advantage of the high dependence of shape on the phase behavior of the final assembled state, in addition to subjecting the system to a boundary condition that has been previously shown to induce rich phase behavior in polygons. The Archimedean truncated tetrahedron was chosen because simulations of truncated tetrahedrons in 3D show rich phase behavior that is highly dependent on the truncation parameter, $t$, (see Supplementary Information) with crystalline structures that are analogous to important atomic crystals. For example, ATTs ($t = 2/3$), which have four regular hexagonal faces and four regular triangular faces with all the same edge lengths, are predicted to form diamond structure at lower packing densities (~0.6), as well as α-arsenic at higher packing densities(~1)[42]. Simulations of truncated polyhedrons (i.e. cubes, octahedrons) constrained to a 2D plane has been previously explored[34], but truncated tetrahedrons have yet to be studied. ATTs have also been studied in simulation under spherical and wall confinements[33], but their 2D behavior on a surface was not further explored. Although the behavior of these types of polyhedrons are of interest, the main experimental limitation is the ability to fabricate such geometries with high monodispersity.

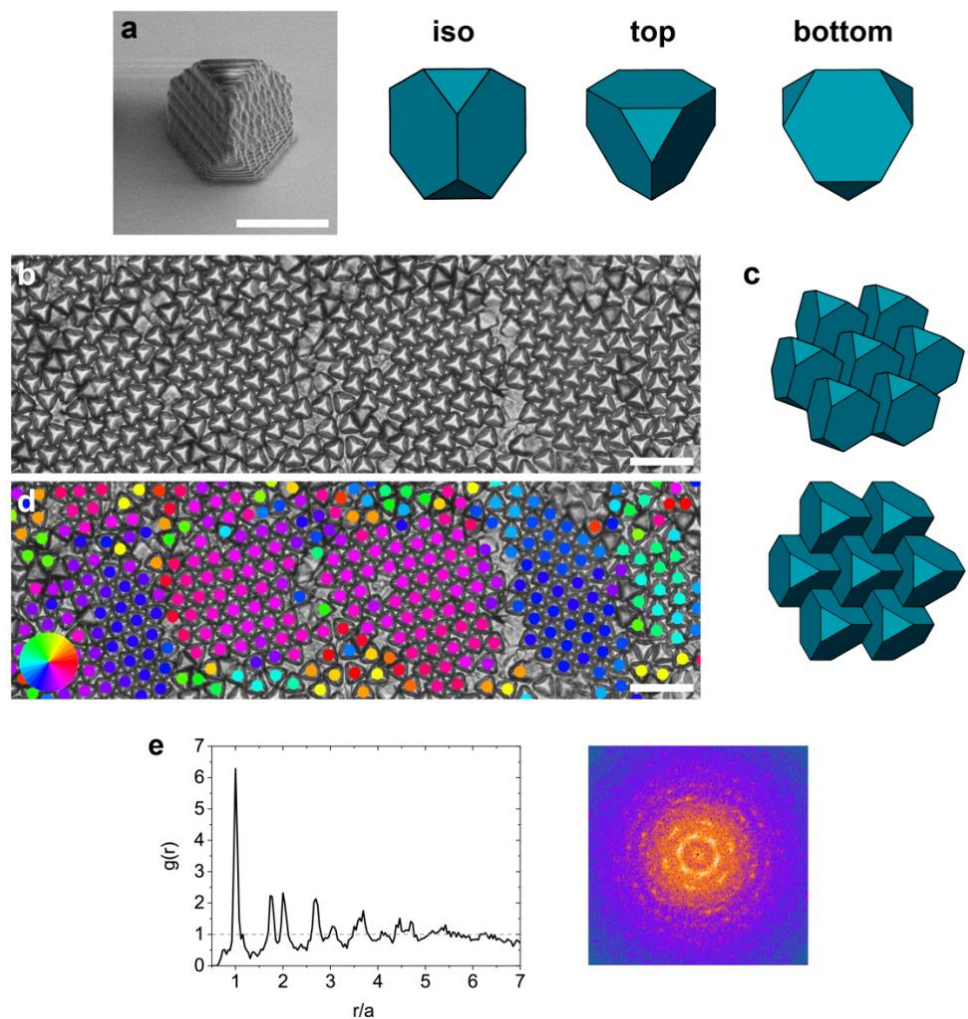

**Fig. 1. Hexatic phase. (a)** SEM image and 3D model of ATT (left to right: isometric, top, bottom view). Scale bar is 5 μm. **(b)** Optical image of self-assembled hexatic structure. Scale bar is 20 μm. **(c)** 3D model of self-assembled structure (isometric and top view). **(d)** Bond orientational order parameter of the particles represented as different colors. Particles with similar colors have similar rotational orientation. Particles with opposite colors on the color wheel are rotated by 30°. Scale bar is 20 μm. **(e)** Pair distribution function, $g(r)$ and Fourier transform of image **(b)**.

To overcome the synthetic challenges of forming polyhedral particles with high monodispersity, two-photon lithography is used to fabricate ATT microparticles with a side length of 3.5 μm **(Fig. 1a)**. Approximately 50,000 particles are fabricated with ≤ 5% variation in particle size[43]. Other tetrahedral particles, such as regular tetrahedrons ($t = 0$) and truncated tetrahedrons ($t = 7/10$), are also easily fabricated using this method (see Supplementary Information). After fabrication, the particles are dispersed in water and deposited in a well plate for assembly. Initially, the particles randomly sediment on the substrate and are dispersed across the substrate with low packing density. We observe that the particles are generally oriented with a hexagonal side facing the substrate, with a triangular face pointing upward, referred to as the 'upright' position. This is due to the center of gravity of the particle being weighed towards the hexagonal face. The substrate is then tilted to apply a small gravitational potential field. After several days (~144 hours), the particles aggregate to one side of the well plate which increases the local packing density and causes ordered regions to form. This also induces a small osmotic pressure that occurs due to particles stacking next to each other.

At a ~5 degree tilt angle, the ATTs form a hexatic phase (**Fig. 1b**). In this phase, the particles, which are oriented with their hexagonal face in contact with the substrate, have 6 nearest neighbors. For this geometry to form, three of the triangular faces of each particle are in face-to-face contact with the hexagonal faces of its neighboring particles as shown in **Fig. 1c**. This effectively "locks" the particle into place by preventing the neighboring particle from moving in the z-direction. To the best of our knowledge, this is the first report of this structure for ATTs in self-assembled colloidal particles under 2D confinement. The grain size and rotational order is analyzed using a bond orientational order parameter that accounts for the 6-fold symmetry of the assembled structures (see Supplementary Information)[42]. This order parameter is represented as colors in **Fig. 1d**. Using this analysis, grains are identified as particles with the same color and found to be ~30 μm or 30-40 particles in size. Grains are separated by vacancies (missing particles) and point defects (disordered particles). The spatial pair distribution function, $g(r)$, is used to quantify the translational packing order (**Fig.**

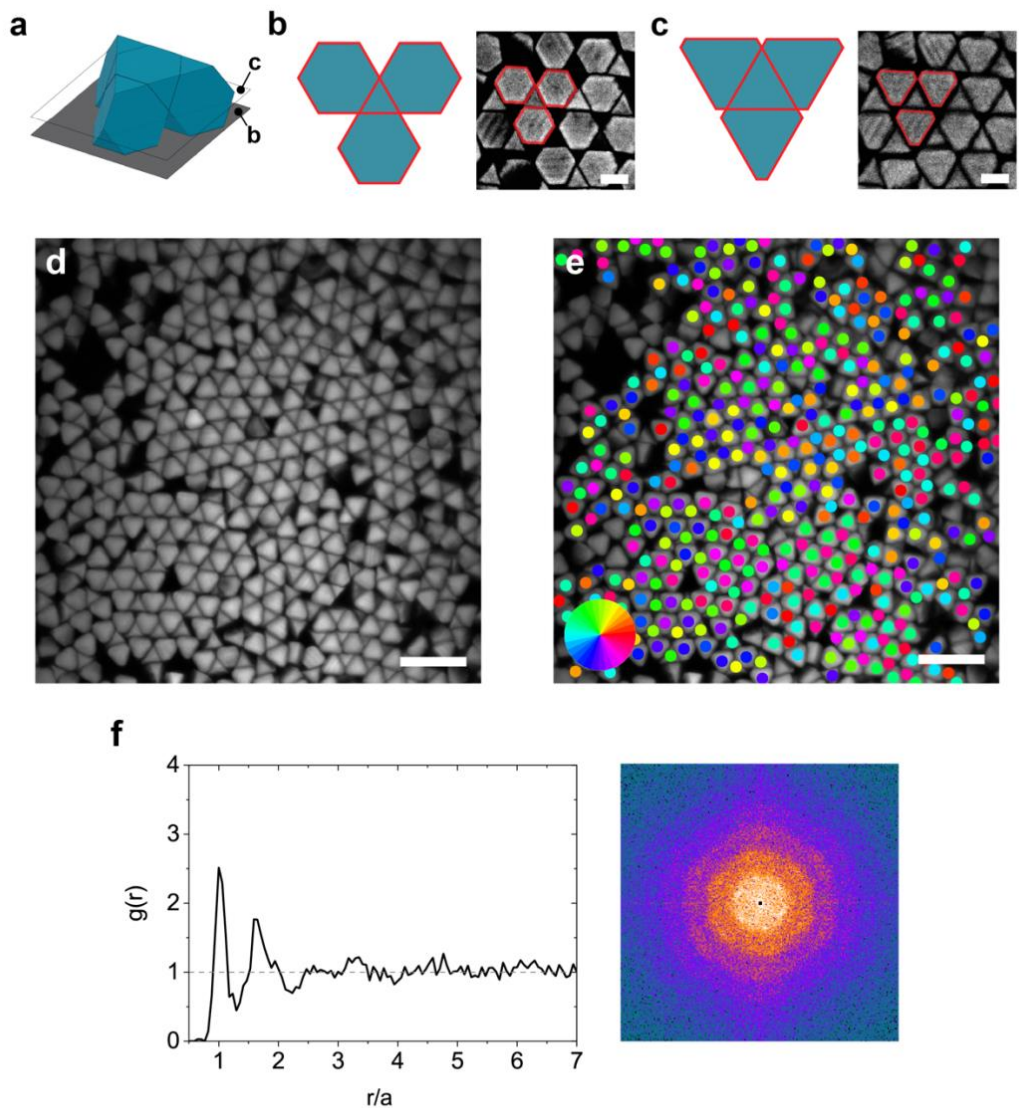

**Fig. 2. Quasi-diamond phase.** (**a**) 3D model of self-assembled structure. The planes that correspond to images b and c are marked. (**b-c**) Confocal images and 2D models of the same quasi-diamond structure at different focal planes. (**b**) is focused at the substrate (gray) and (**c**) is focused at the middle of the particle. The red outline shows the analogous geometry between the model and the confocal images. Scale bars are 5 μm. (**d**) Confocal image of a large region of the sample. Scale bar is 20 μm. (**e**) The bond orientational order parameter of the particles is represented as different colors. Adjacent particles with opposite colors on the color wheel indicate the quasi-diamond structure (e.g. green and pink). Scale bar is 20 μm. (**f**) Pair distribution function, g(r), and Fourier transform of image (**d**).

**1e**). The *g(r)* plot shows a first peak at ~6.6 μm followed by a double peak. This is indicative of a hexatic phase, which has been observed in 2D assemblies of spherical colloids on a surface[44]. The corresponding Fourier transform shows bright spots in a hexagonal geometry, which is indicative of a hexatic phase.

The ATTs are then tilted by an additional ~5 degrees to an angle of ~10 degrees and allowed to assemble over 48 hours. This results in a phase that is drastically different than the previous hexatic phase, with triatic cluster phases. This is reflected in the optical images as triangular shapes that are arranged with 3 nearest neighbors. Because of the drastically different particle shape in the optical microscope, the particle orientations are elucidated by using confocal imaging at different z-planes (**Fig. 2a-c**). These images show alternating triangular and hexagonal faces near the substrate. As we focus away from the substrate, the triangular faces become larger, and the hexagonal faces become more triangular (**Fig. 2b-c**). This demonstrates that the ATTs form a nearly space-filling structure made up of a two-particle unit cell that consists of one 'upright' facing ATT and one 'upside-down' facing ATT. This structure, which we refer to as quasi-diamond, is equivalent to a two-atom basis in diamond cubic structure. This has been predicted to be the lowest energy 3D structure of ATTs that self-assemble under an entropic driving force[42].

The bond orientational order parameter is calculated for the quasi-diamond structure (**Fig. 2e**). These images are obtained at a focal plane near the center of the particle which cause the ATTs to appear as triangular shapes under these imaging conditions. Grains are shown as regions of alternating colors and are approximately half the size (~20 um) of the hexatic grains. The spatial pair distribution function, $g(r)$, and corresponding Fourier transform show a first peak at ~4.4 um, a weaker second peak, and no additional peaks (**Fig. 2f**). This indicates the formation of a 3-fold symmetric phase with short range order, small grains, and a higher density of defects as compared to the hexatic phase. This $g(r)$ and bond orientational order is similar to that of self-assembled triangular plates that form 3-fold, 2D structures[14], as well as self-assembled, two-photon lithographed regular tetrahedrons (see Supplementary Information).

Here, we consider the thermodynamics of self-assembly. For hard particle systems, this behavior can be examined through the lens of entropy maximization[45]. In these systems, self-assembly is dominated by an entropic driving force due to the gain in free volume when the particles form an ordered arrangement. Generally, the free volume is maximized when particles are in face-to-face arrangements[46]. For the disordered (right after deposition) to hexatic phase transition, the increase in face-to-face area is due to the contact of the three of the triangular faces of the ATT with the hexagonal faces of its neighboring particles. The hexatic to quasi-diamond phase transition results in further gains in entropy because the face-to-face contact increases by >100%. The free volume change between hexatic and quasi-diamond phase can also be computed directly. The total volume of the system is considered as an x-y box that fits $N$ particles, with a z-height of one particle unit. Using this as the total volume accessible to the particles, the hexatic phase has a maximum packing fraction of ~64% while the quasi-diamond structure is a nearly space-filling structure at ~99%. Therefore, the relative change in volume density between hexatic phase and the quasi-diamond phase is ~50%. This indicates that there is a large driving force towards the quasi-diamond phase from the hexatic phase. Other truncated tetrahedrons ($t$ = 7/10) form even smaller quasi-diamond grains because of a lower change in free volume and lower driving force for self-assembly (see Supplementary Information).

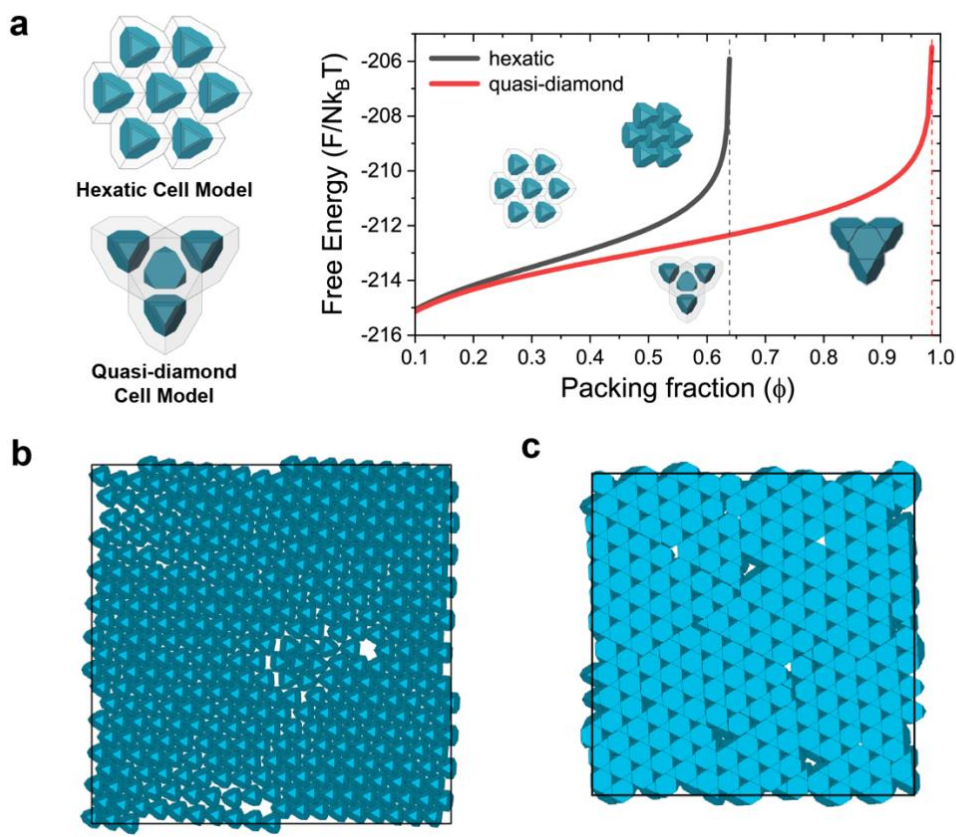

**Fig. 3. Analytical model and hard particle Monte Carlo simulation. (a)** Geometric models of self-assembled particles within volumetric cells and the resulting free energy calculations. **(b)** Monte Carlo simulation of ATTs constrained to a 2D plane and laterally compressed results in the formation of a hexatic phase. **(c)** Monte Carlo simulation of ATTs after removal of the 2D constraint. Continued lateral compression leads to the formation of the quasi-diamond phase.

An approximate single-cell occupancy model is used to estimate free energy ($F$) as a function of packing fraction ($\phi$) (**Fig. 3a**)[47,48]. This model is one of the only methods to analytically calculate the free energy of a hard-particle system and has only been used to model hard spheres in different phases[49–53]. The model uses the phase of interest (for spheres, face-centered cubic or hexagonal closed-packed) at the highest packing fraction and partitions each particle center inside Voronoi polyhedrons, known as cells. For spheres packed in a face-centered cubic or hexagonal closed-packed phase, the corresponding Voronoi polyhedron would be a dodecahedron. This model assumes that each particle is constrained within its own cell and can only access the volume associated with its own cell. The Voronoi cell can then be scaled equally in 3D to decrease the packing fraction, and effectively increase the volume accessible to each particle. The total accessible volume of each cell can then be used to estimate the free energy of the system as a function of packing fraction, $\phi$:

$$F(\phi)/Nk_BT \sim ln(V_{free}(\phi)) \qquad (1)$$

The single-cell model is most accurate at higher packing fractions when the assumptions are more likely to be satisfied. For lower packing fractions, this error is associated with a "communal entropy"[52]. For our system, which consists of polyhedron shapes and is quasi-2D, a slightly different procedure is taken. Instead of constructing Voronoi cells and dilating in 3D, self-similar cells are constructed around each particle from their phase (either hexatic or quasi-diamond) and then dilated in two dimensions (**Fig. 3a**). The accessible free volume, $V_{free}$, is then taken as $V_{cell} - V_{ATT}$.

We find that the quasi-diamond phase has a slightly lower free energy than the hexatic phase at all packing fractions. However, this difference becomes larger when packing fraction increases, especially when it approaches the maximum hexatic packing density. As the packing fraction reaches the theoretical maximum packing fraction of its phase, the free energy goes asymptotically to infinity as this corresponds to hard particle collisions or overlaps. Although the hexatic phase approaches this limit at a packing fraction of ~0.64, the quasi-diamond phase is not accessible until there is sufficient energy to overcome the thermodynamic or kinetic barrier of this transition. We attribute this barrier to the effects of the quasi-2D confinement, which prevents particle out-of-plane rotation. A thermodynamic barrier exists due to the gravitational potential. A hexatic to quasi-diamond phase transition would require 50% of the particles to flip from an 'upright' orientation to an 'upside-down' orientation. This is associated with a potential energy change, due to the increase in height of the center of mass. A kinetic barrier also exists due to the free volume required to rotate a particle out-of-plane.

We can test the hypothesis that the phase transition energy barrier is related to an out-of-plane particle rotation by using hard particle Monte Carlo simulations (**Fig. 3b-c**) (also see Movie S1). ATTs are first confined to a 2D plane such that the particles cannot rotate out-of-plane and can only move in the x-y directions. The particles are laterally compressed until they approach the maximum theoretical packing fraction. This leads to the formation of an hexatic structure, with some defects (**Fig. 3b**) as seen in experiments. This 2D constraint is then removed, which allows out-of-plane rotation. Once this constraint is removed, the particles almost immediately form the quasi-diamond phase under continued lateral compression (**Fig. 3c**).

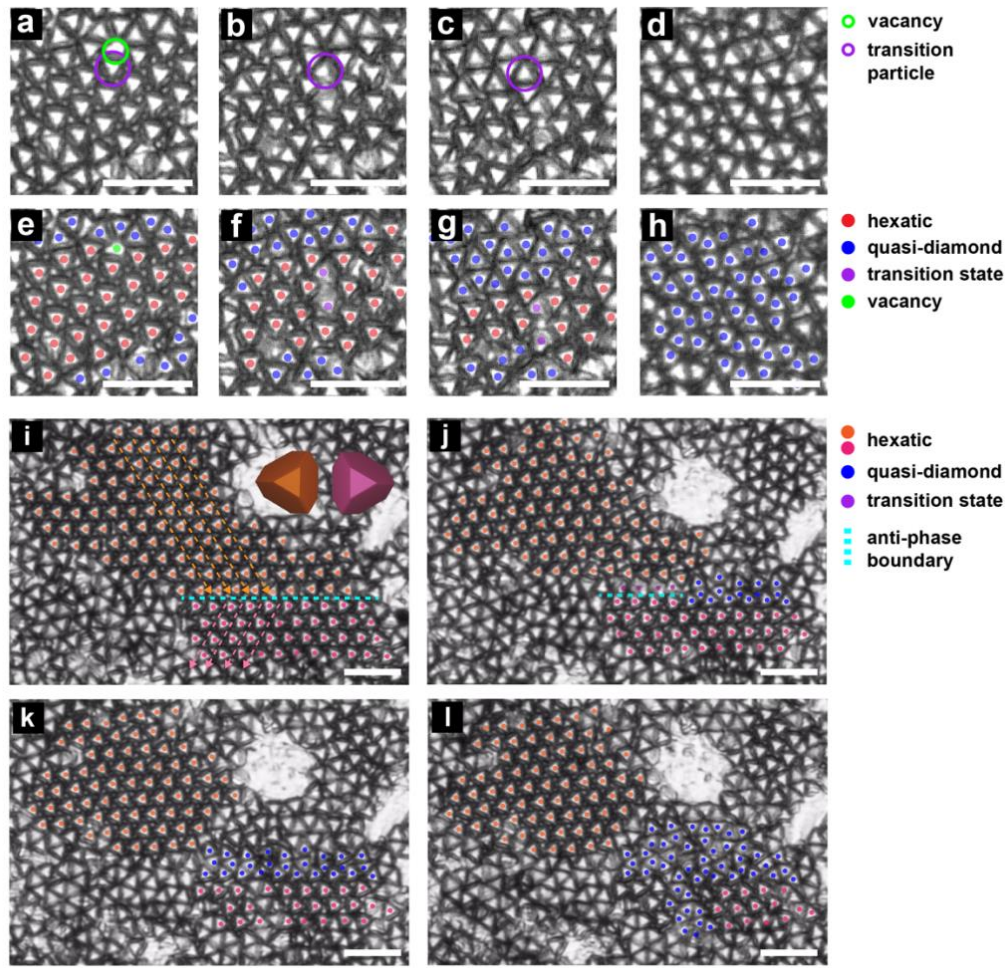

**Fig. 4. Direct imaging of defect mediated phase transitions.** In-situ optical images of the **(a)** initial hexatic grain and vacancy, **(b)** first particle rotation, **(c)** propagation of the phase transition through the hexatic grain, **(d)** final quasi-diamond state. The vacancy is marked by a green circle. The adjacent ATT is marked by a purple circle, which is the first particle to transform. **(e-h)** The same images with colors that indicate the hexatic phase (red), quasi-diamond phase (blue), particle in transition (purple), and vacancy (green). **(i)** Two hexatic grains with different orientations are shown in orange and pink with corresponding 3D models. These grains are separated by an anti-phase boundary (blue dashed line). Orange and pink arrows show the alignment of the particles and point in the direction of a triangular vertex. **(j)** Transition of hexatic grains (orange or pink) to quasi-diamond (blue) at the anti-phase boundary is preceded by the rotation of particles into a transition state (purple) along these rows. **(k)** The anti-phase boundary is replaced by the quasi-diamond phase (blue) which separates the two remaining hexatic grains (orange or pink). **(l)** The phase transition begins to propagate in the lower grain and transform the hexatic phase (pink) to the quasi-diamond phase (blue). All scale bars are 25 μm.

In-situ optical microscopy reveals the kinetics of phase transition. First, a hexatic sample is assembled through a small tilt angle (~ 5 degrees) as previously described. This sample is then tilted by an additional ~5-10 degrees and moved to the microscope stage. By the time that imaging begins (~10 min after tilting to ~5-10 degrees), many hexatic regions have already transformed to quasi-diamond. However, the transition of the remaining hexatic phase can be observed.

These in-situ experiments show that the phase transition is mediated by defects and that these defects allow for out-of-plane particle rotation. **Fig. 4** shows specific instances of these defect induced phase transitions (see Move S2 and S3). **Fig. 4a-h** show a vacancy mediated phase transition. Initially, a hexatic grain is surrounded by the quasi-diamond phase with a vacancy present near the phase boundary. The hexatic particle adjacent to the vacancy rotates out-of-plane and is upside down, with its triangular face facing the substrate. This leads to a chain reaction in which the next particle rotates and transforms, and then the next particle, until the hexatic phase has fully transformed into a quasi-diamond phase. The presence of the vacancy seems to facilitate an out-of-plane rotation of the ATT particle by providing the free volume to accommodate an out-of-plane rotation.

Direct observation of a phase transition is also observed at an anti-phase boundary between two hexatic grains which have particles oriented in different directions (**Fig. 4i-l**). To the best of our knowledge, this is the first report of this type of phase boundary for self-assembled ATTs. The two rows above (orange hexatic grain) and the row below (pink hexatic grain) the anti-grain boundary (dashed blue line) undergo a phase transition to the quasi-diamond phase. The transition occurs rapidly for half the particles, while the remaining particles in these rows begin to rotate into a transition state (begin flipping out-of-plane) (**Fig. 4j**). This is followed by the transition of the remaining particles at the anti-phase boundary into the quasi-diamond phase (**Fig. 4k**) and further growth of the quasi-diamond phase until two smaller, isolated hexatic grains remain (**Fig. 4l**).

Without these defects, the hexatic to quasi-diamond phase transitions is kinetically improbable: an ATT particle would need to escape from its "locked" hexatic configuration, and then rotate out-of-plane. This kinetic pathway is unlikely, given that the "locked" hexatic configuration geometrically prevents out-of-plane motion. However, once a particle successfully rotates into a quasi-diamond phase, the local packing density of the particles around it is lowered because the quasi-diamond phase is ~50% denser than the hexatic phase. This allows neighboring particles to also have more free space to rotate out-of-plane and continue propagating the phase transition. This type of defect mediated transition is also seen in the simulations, right after the removal of the 2D constraint (see Movie S1). By analyzing a hexatic to quasi-diamond phase transition, the phase transition rate was found to follow Avrami's solid-solid phase kinetic theory in 2D[54,55] (see Supplementary Information).

In summary, we have assembled Archimedean truncated tetrahedrons under quasi-2D confinement and shown a hexatic phase that has not been previously reported in literature for this shape. We directly imaged a novel phase transition from a hexatic phase, which has 6 nearest neighbors, into a quasi-diamond phase, which has 3 nearest neighbors. We determined the thermodynamics and kinetic mechanism of this phase transition using analytical and computational methods. Other 3D polyhedral geometries can be easily fabricated using 3D nanoprinting methods, such as two-photon lithography, to access a huge phase space of additional crystal phases, especially when under quasi-2D confinement. While the size of the current lattices is too large for optical frequency photonic crystals or metamaterials, two-photon lithographed structures can be shrunk up to ~20% of their original size to form sub-micron scale particles through pyrolysis[56]. In addition, chemistries exist for directly printing high dielectric materials such as silica, which is also necessary for optical applications[57]. Magnetic, plasmonic and luminescent nanoparticles can be incorporated into photoresists to impart further functionality and enable self-assembly under external stimuli. This could be used to generate a novel class of programmable matter in which dynamic phase transitions are used to switch between structures and properties.

## Acknowledgements

We thank Prof. Matthew Jones for helpful advice on experiments and the manuscript. DD acknowledges the National Science Foundation Graduate Research Fellowship under Grant No. 1656518. JK is supported by a Stanford Graduate Fellowship. DD, JK, and XWG acknowledge funding from the Hellman Foundation, and the National Science Foundation under Grant No. CMMI-2052251. Part of this work was performed at the Stanford Nano Shared Facilities (SNSF), which is supported by the National Science Foundation under award ECCS-1542152. Part of this work was performed at the Stanford Cell Sciences Imaging Facility.